\documentclass[12pt,preprint]{aastex}
\usepackage{subfigure}

\def\lya{\ifmmode {{\rm Ly}\alpha}\else
	Lyman-$\alpha$\fi}

\def\f17{f_{17}}

\def\kpc{\hbox{kpc}}

\def\arcsec{\ifmmode {''}\else{$''$}\fi}
\def\sqarcsec{\ifmmode{\square''}\else{$\square''$}\fi}

\def\kJy{\hbox{kJy}}

\def\ergcm2s{\ifmmode {\rm\,erg\,cm^{-2}\,s^{-1}}\else
                ${\rm\,ergs\,cm^{-2}\,s^{-1}}$\fi}
\def\ergsec{\ifmmode {\rm\,erg\,s^{-1}}\else
                ${\rm\,ergs\,s^{-1}}$\fi}
\def\kmsMpc{\ifmmode {\rm\,km\,s^{-1}\,Mpc^{-1}}\else
                ${\rm\,km\,s^{-1}\,Mpc^{-1}}$\fi}

\def\oiiipair{[\ion{O}{3}] $\lambda \lambda$4959,5007}

\begin{document}

\title{Spectroscopic Confirmation of Faint Lyman Break Galaxies at Redshifts
Four and Five in the Hubble Ultra Deep Field}

\author{
James E. Rhoads\altaffilmark{1,9},
Sangeeta Malhotra\altaffilmark{1},
Norbert Pirzkal\altaffilmark{2},
Mark Dickinson\altaffilmark{3},
Seth Cohen\altaffilmark{1},
Norman Grogin\altaffilmark{1,2},
Nimish Hathi\altaffilmark{1},
Chun Xu\altaffilmark{2},  
and the PEARS team
}

\begin{abstract}
We present the faintest spectroscopically confirmed
sample of $z\sim 5$ Lyman break galaxies to date.  
The sample is based on slitless grism spectra of the Hubble
Ultra Deep Field region from the GRAPES
(Grism ACS Program for Extragalactic Science) 
and PEARS (Probing Evolution and Reionization Spectroscopically)
projects, using the G800L grism on the HST 
Advanced Camera for Surveys.  We report here confirmations
of 39 galaxies, pre-selected as candidate Lyman break
galaxies using photometric selection criteria.  We compare
a ``traditional'' V-dropout selection to a more liberal one
(with $V-i > 0.9$), and find that the traditional criteria
are about $64\%$ complete and $81\%$ reliable.
We  also study the \lya\ emission properties of our
sample.  We find that \lya\ emission is detected in 
$\sim 1/4$ of the sample, and that our broad-band
color selected sample includes $\sim 55\%$ of previously 
published line-selected \lya\ sources.
Finally, we examine our stacked 2D spectra.  We demonstrate
that strong, spatially extended ($\sim 1''$)  \lya\ emission 
is not a generic property of these Lyman break galaxies,
but that a modest extension of the \lya\ photosphere (compared
to the starlight) may be present in those galaxies with
prominent \lya\ emission.
\end{abstract}


\altaffiltext{1}{Arizona State University}
\altaffiltext{2}{Space Telescope Science Institute}
\altaffiltext{3}{National Optical Astronomy Observatory}
\altaffiltext{9}{Email: James.Rhoads@asu.edu}

\section{Introduction}
Star forming galaxies in the early universe have been found in large
numbers both by looking for strong Lyman breaks, and by looking for
\lya\ line emission.  Galaxies found by these two methods differ
greatly in their typical properties.
This may indicate physically distinct galaxy populations, or
selection effects inherent in the search methods, or a combination of
the two.  To help address these issues, we here examine the selection
of Lyman break galaxies in the Hubble Ultra Deep Field (HUDF).
The depth of the HUDF images means that  \lya\ emitting
galaxies with fluxes typical of present surveys should all be
detected (down to flux $10^{-17} \ergcm2s$), even if they have 
no continuum emission at all.  We combine these deep HUDF images
with the deepest slitless spectra ever obtained, from the GRAPES
and PEARS projects (see below).  
These slitless spectra allow us to look for prominent
\lya\ line emission, with or without pre-selection for a Lyman
break.  We study the continuum properties of a set of \lya\ 
selected galaxies, to see what fraction pass our Lyman break
crieteria.   Conversely, we also study the emission line properties
of a Lyman break selected sample.  Moreover, we examine 
spectra for Lyman break samples selected with two sets
of  photometric criteria, one ``traditional'' and the other
more inclusive.

The Grism ACS Program for Extragalactic Science (GRAPES) project and
Probing Evolution And Reionization Spectroscopically (PEARS) project 
are slitless spectroscopic surveys that exploit the potential of the G800L
grism on the Hubble Space Telescope's Advanced Camera for Surveys
(ACS) to achieve the most sensitive unbiased spectroscopy yet at
red optical wavelengths ($0.58 \mu m \la \lambda \la 0.96 \mu m$).  
Two primary factors enhance our sensitivity relative to ground-based
spectrographs.  First, high redshift galaxies are typically compact
(e.g., Ferguson et al 2004; Pirzkal et al 2007; 
Hathi, Malhotra, \& Rhoads 2008), so that the sensitivity of ground-based
observations is hampered by atmospheric blurring of the galaxy images
(and also usually by slit losses).  Second, the OH emission line
forest in the night sky spectrum raises the background level for
ground-based observations, introducing random noise (and often
systematic residuals as well) in red-light spectra of high redshift
galaxies from the ground.  Our HST spectra are free of both effects.
Moreover, because the ACS grism is a slitless spectrograph, we obtain
spectra of every source in our field of view, with no preselection
required.

GRAPES was targeted in the HUDF region, to
complement the HUDF direct images, which are in turn the deepest
optical imaging to date (Beckwith et al. 2006).  The GRAPES survey,
and in particular our data analysis methods, are described in more
detail by Pirzkal et al. (2004).  PEARS (further described in 
Malhotra et al 2008) included an additional forty orbits of G800L 
integration on the HUDF, essentially doubling the integration time.
A primary scientific goal of both surveys is to study the
properties of Lyman break galaxies using spectroscopically
confirmed samples at unprecedented sensitivity.
We are pursuing this effort through a targeted look at
Lyman break candidates identified in the HUDF using both
an $i$ dropout criterion (Malhotra et al 2005) and a $V$ 
dropout criterion (this paper).  

We refer to ACS filters by names of roughly corresponding
ground-based  filters: F435W $\rightarrow$ B; F606W $\rightarrow$ V;
F775W $\rightarrow$ i; and F850LP $\rightarrow$ z.
Throughout this paper we use  the current concordance
cosmology ($H_0 = 71 \kmsMpc$, $\Omega_M = 0.27$, $\Omega_{total} = 1$;
see Spergel et al. 2003, 2007).  We use AB magnitudes,
so that magnitude zero corresponds to $f_\nu = 3.6 \kJy
= 3.6 \times 10^{-20} \ergcm2s {\hbox{Hz}^{-1}}$.

\section{The Samples} \label{sample}
We consider both Lyman break and \lya\ emission selected samples.  
For \lya\ selection, we use the previously published sample
from Pirzkal et al (2007; hereafter P07), which in turn is
based on the HUDF emission line catalog of Xu et al (2007).
This sample contains 9 \lya-emitting galaxies, spanning
redshifts $4.0 \le z \le 5.76$, line fluxes
$2 \la f/10^{-17} \ergcm2s \la 6$, and continuum 
magnitudes $25.5 \la i \la 29$.

For Lyman break selection, we derive new samples based on 
HUDF photometric pre-selection combined with GRAPES/PEARS grism 
spectroscopy.  
Because we have spectra for all objects in our sample,
we can examine a more inclusive set of Lyman break 
galaxy candidates than is practical for photometric studies.
We examined as candidate LBGs all objects having $V-i > 0.9$,
and  $i<27.7$.  The magnitude cut prevents the sample from 
being swamped by galaxies too faint for accurate photometry or 
spectroscopy.  We inspected visually the GRAPES and PEARS
spectra of all objects passing these selection cuts, and
retained in our final sample those objects whose spectra
support a Lyman break identification.

\begin{deluxetable}{lllllrlll}
\tablecolumns{9}
\tablewidth{0pc}
\tablecaption{Properties of GRAPES/PEARS HUDF V-dropout Sample}
\tablehead{
\colhead{Object ID} & \colhead{RA} & \colhead{Dec} &
 \colhead{$i$ mag}  & \colhead{$V-i$} &
  \colhead{$i-z$} &\colhead{$B$} & \colhead{Redshift} & \colhead{Grade}
\\}
\startdata
119 & 53.1660037 & -27.8238734 & 27.51 & 2.14 & -0.62 & 28.97 & 4.88 & 2.5 \\
201 & 53.1649949 & -27.8224088 & 27.59 & 1.23 & -0.24 & 32.92 & 4.60 & 2 \\
478 & 53.1733579 & -27.8182711 & 27.06 & 1.11 & -0.06 & 30.65 & 4.52 & 2 \\
546 & 53.158038 & -27.8179411 & 27.86 & 3.13 & 0.79 & 31.52 & 5.42 & 2.5 \\
577 & 53.1614532 & -27.8174379 & 27.08 & 1.01 & 0.09 & 30.55 & 3.8 & 2.5 \\
646 & 53.1660559 & -27.8167778 & 27.44 & 2.22 & 0.24 & 30.18 & 4.9 & 2 \\
712 & 53.1783554 & -27.8162519 & 27.26 & 2.32 & 0.13 & 32.4 & 5.12 & 1.8 \\
1115 & 53.1722726 & -27.8119732 & 26.40 & 1.85 &  0.17 &  28.89 & 4.72 & 1.8 \\
1392 & 53.1563588 & -27.8095883 & 27.75 & 3.96 & -1.05 & 30.75 & 4.90 & 1 \\
2408 & 53.1885357 & -27.8034637 & 26.76 & 1.77 & 0.14 & 31.82 & 4.86 & 2 \\
2599 & 53.1626689 & -27.8022982 & 27.00 & 1.76 & -0.07 & 29.75 & 4.88 & 2 \\
2736 & 53.1499325 & -27.8017358 & 26.94 & 1.39 & 0.11 & 32.27 & 4.56 & 2 \\
2881 & 53.1415872 & -27.8005681 & 25.84 & 1.78 & 0.18 & 29.79 & 4.56 & 2 \\
2894 & 53.1462482 & -27.8008077 & 27.58 & 2.57 & -0.10 & 29.85 & 5.3 & 2 \\
2898 & 53.1798203 & -27.8008748 & 27.06 & 1.79 & -0.32 & 29.50 & 4.67 & 2.3 \\
3094 & 53.1514271 & -27.7997637 & 25.72 & 1.18 & 0.02 & 29.57 & 4.62 & 1 \\
3250 & 53.1326676 & -27.7989430 & 27.14 & 3.74 &-0.29 &30.12 & 4.90 & 2.3 \\
3968 & 53.1833331 & -27.7959556 & 27.56 & 3.35 &-0.82 &31.55 & 4.7 & 2 \\
5183 & 53.1437786 & -27.7908654 & 27.38 & 0.91 & -0.57 & 32.92 & 4.62 & 1 \\
5225 & 53.1385743 & -27.7902115 & 25.83 & 1.52 & 0.04 & 27.87 & 5.42 & 2 \\
5296 & 53.1907601 & -27.7903482 & 27.08 & 2.71 & 0.29 & 31.78 & 5.14 & 2.2 \\
5307 & 53.1908538 & -27.7903656 & 27.15 & 2.53 & 0.38 & 30.39 & 5.14 & 2.2 \\
5788 & 53.1456573 & -27.7882186 & 27.46 & 4.32 & 0.50 & 30.27 & 5.1 & 2 \\
6066 & 53.1845519 & -27.7869713 & 26.11 & 0.93 & 0.01 & 29.12 & 4.42 & 1 \\
6139 & 53.1581263 & -27.7863866 & 25.49 & 1.37 & -0.11& 31.51 & 4.68 & 1.2 \\
6515 & 53.1273697 & -27.7851656 & 27.15 & 1.25 & -0.24 &  29.75 & 4.75 & 1 \\
6681 & 53.1926615 & -27.7841483 & 27.10 & 2.25 & 0.41 & 28.81 & 5.08 & 1.3 \\
7050 & 53.1510392 & -27.7828658 & 27.36 & 4.13 & 0.50 & 30.44 & 5.45 & 2.2 \\
7352 & 53.1376954 & -27.7812680 & 26.87 & 2.88 & -0.02 & 29.37 & 5.04 & 2.5 \\
8301 & 53.1671692 & -27.7745246 & 27.18 & 2.91 &  0.22 & 32.47 & 5.0 & 2 \\
8664 & 53.1890652 & -27.7770042 & 26.77 & 2.08 & 0.08 & 29.9 & 4.9 & 1.3 \\
8682 & 53.1888007 & -27.7770931 & 25.62 & 1.94 & 0.06 & 28.70 & 5.08 & 2 \\
8896 & 53.1900052 & -27.7790544 & 26.92 & 1.56 & 0.09 & 31.30 & 4.33 & 2 \\
9040 & 53.1711852 & -27.7784585 & 25.97 & 3.42 & -0.49 & 28.3 & 4.55 & 2 \\
9057 & 53.1829957 & -27.7804592 & 26.90 & 0.91 & -0.02 & 31.84 & 4.17 & 2 \\
9275 & 53.1531485 & -27.766181 & 25.83 & 1.26 & 0.14 & 28.93 & 5.12 & 2 \\
9777 & 53.1702387 & -27.7628552 & 26.17 & 1.95 & 0.85 & 30.20 & 5.41 & 2.5 \\
9983 & 53.1671627 & -27.7598546 & 25.62 & 1.79 & 0.17 & 32.27 & 4.82 & 2 \\
20191 & 53.1725558 & -27.8137100 & 25.73 & 1.41 & 0.09 & 31.48 & 4.62 & 1 \\
\enddata
\tablecomments{
Properties of V dropout sources from GRAPES + UDF data.  ``ID'' is the
HUDF catalog number, since we used the HUDF catalog as the master
object list for GRAPES.  The ``Grade''
column indicates the quality of a candidate as assessed by visual inspection
of the GRAPES and PEARS spectra.  
Objects assigned grade 1 are very secure Lyman
break galaxies, while those assigned grade 2 are probable LBGs.  Some objects
are assigned non-integer grades between these possibilities.
We also identified ``grade 3'' objects, for which
our data could neither confirm nor refute a photometrically identified
Lyman break candidate, and ``rejected'' objects, shown not to be
Lyman break objects by their grism spectra.  We do not tabulate these
lower category sources.
\label{vdrop_tab}}
\end{deluxetable}

To determine redshifts, we fitted our spectra with
a model consisting of a power law continuum attenuated by \lya\ forest
absorption, which we modeled using the formalism of Madau (1995).
We thus fitted three parameters to each spectrum: the flux normalization,
spectral slope, and redshift.  In cases with strong \lya\ line
emission, the fitted slope is strongly biased towards blue values,
since we do not expliclitly fit emission lines.

We then assessed the best fitting model for each spectrum and assigned
a grade to each.  These grades were based on the
$\chi^2$ of the fit, the signal to noise in the spectrum, and a visual
inspection of the fit.  Grade~1 was given to the highest quality fits,
where the LBG identification and redshift are very secure.  Grade~2
was given to reasonably secure Lyman break objects.  Grade~3 was given to
sources that remain possible Lyman break objects but that cannot be
confirmed, whether because the signal-to-noise was insufficient,
or because the spectra suffered from contamination. Finally, grade~4 was
assigned to photometric candidates that were refuted by the grism spectra.
We assigned fractional grades to a modest number of galaxies where it 
seemed warranted.

\begin{figure}
\plotone{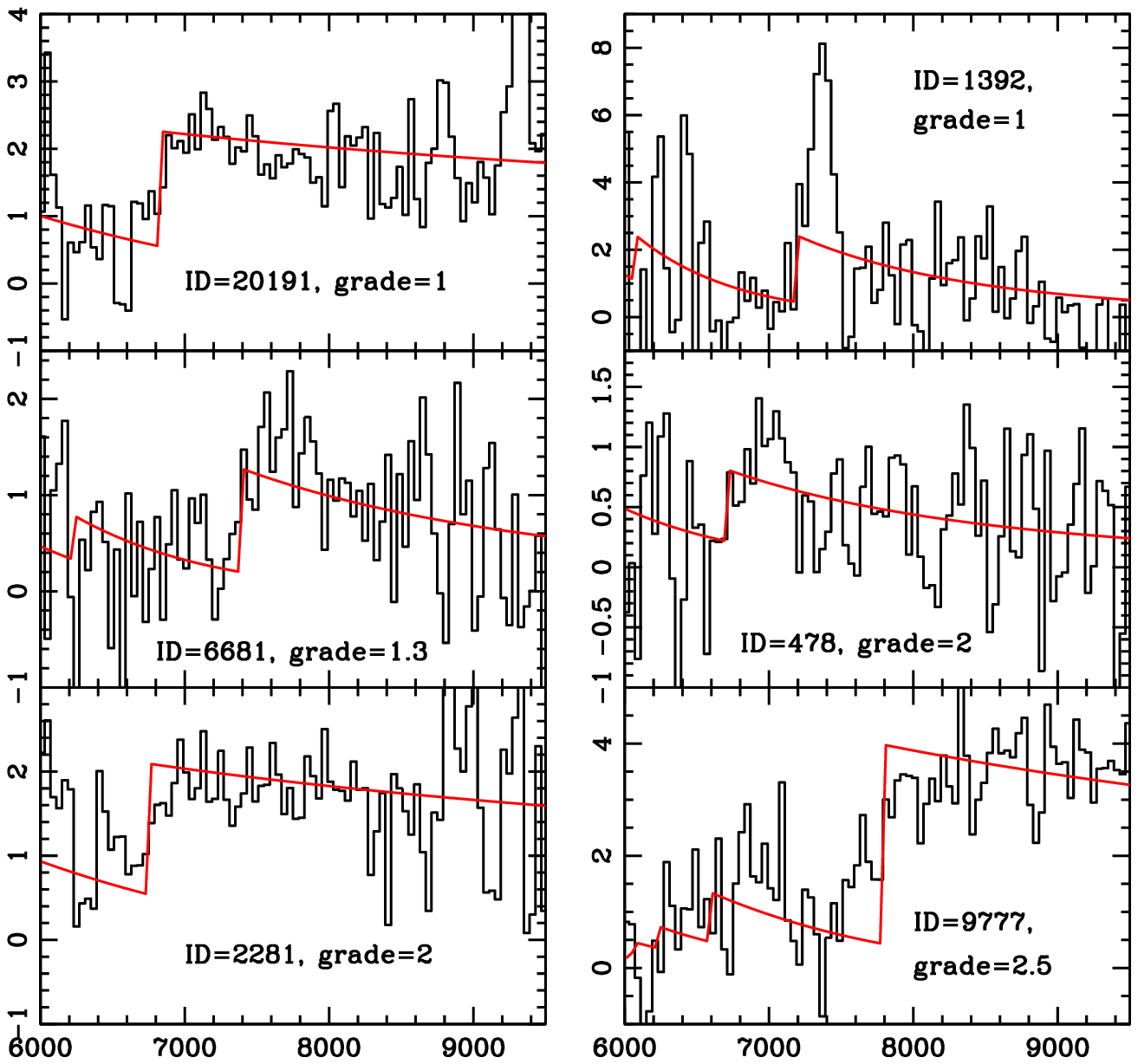}
\caption{Six representative spectra from our HUDF V dropout sample.
These span a range of quality, from grade 1 (best) to grade 2.5
(marginal).  The grades were assigned by examination of both the
1D spectra (shown here after coadding all position angles)
and the 2D spectra, in both cases examining both results from 
individual position angles and their average.  The best fitting
power law spectrum, attenuated by the IGM opacity based on Madau 1995,
is overplotted.  One of the plotted objects (1392) has a \lya\ line,
which is not included in the fitted model but which was used to establish
the object's redshift.
\label{fig:examples}
}
\end{figure}

In total, we considered 216 candidates fulfilling the photometric
selection criteria.  A few additional objects were examined in the
early stages of the project, resulting in the inclusion in the final sample
of two V dropouts (IDs 546 and 1392) having $27.7<i<27.9$.
In total, we classified 39 objects as ``good'' (1 $\le$ 
grade $\le$ 2.5)
Lyman break objects, 86 as refuted (grade~4) objects, and 
94 as grade~3 (``unidentified'').
Representative spectra for grades from 1 to 2.5 are shown in
figure~\ref{fig:examples}.
The brightest confirmed Lyman break
galaxies have $i \approx 25.5$.  The fraction of grade~3 (``unidentified'')
sources rises steadily from 0 at $i < 25$ to 100\% at $i \approx 27.7$.  
A very rough linear fit is $f \approx 0.3 + 0.2\times (i-26)$.
Among those sources adequately classified using the grism spectra,
the fraction confirmed as Lyman break galaxies {\it rises} weakly 
towards fainter magnitudes, from $\sim 25\%$ at $i\approx 25.5$
to $\sim 50\%$ at $i\approx 27.5$.  To understand this, consider
the primary contaminants of the photometrically selected
Lyman break candidate lists.  These are red stars in our Galaxy, 
and early type galaxies at intermediate redshifts.  While the number-magnitude
relation for Lyman break galaxies is steeply rising around $i \approx 
26$, the number-magnitude relations for both contaminants are much
flatter, and therefore the LBG fraction rises at the faint end.

The V dropout sample includes a range of morphologies, from simple, 
compact galaxies to extended sources 
with tails and/or multiple peaks (e.g., HUDF~5225; Rhoads et
al 2005).
Detailed  morphological analysis of these and other high redshift
UDF sources has been discussed by Pirzkal et al (2005, 2006).

\section{Discussion} \label{disc}

\begin{figure}
\epsscale{0.8}
\plotone{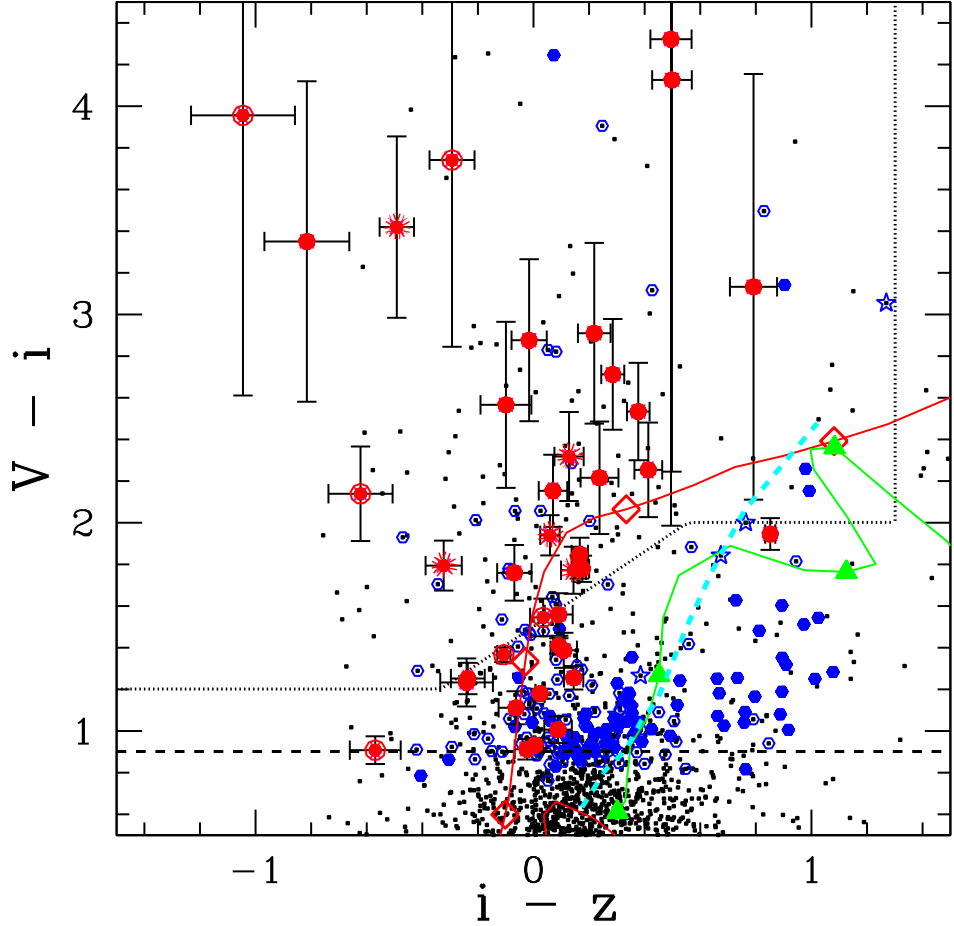}
\caption{The color-color plane for V dropout selection.
Small black dots show all objects in the parent catalog 
having $i<29$ and i- and z-band magnitude errors $<0.3$ mag.  
We examined the spectra of those objects with 
$V-i > 0.9$, $i<27.7$, and $B>27$.  Those confirmed
as Lyman break galaxies are shown as red circles with
$1\sigma$ error bars.  Among the confirmed sources, those
with confirmed \lya\ emission are circled, and those with suspected
\lya\ emission are marked by projecting ``rays.'' 
Objects that were firmly rejected from the Lyman break sample
(i.e., grade 4 objects) are shown as filled blue dots.  
Stars (from the list of Pirzkal
et al 2005) are shown by open blue star symbols.  Finally, 
candidates for which we have insufficient basis for judgment are
shown as open blue circles.  Candidates below the
$V-i=0.9$ line all lie within $1\sigma$ of that cutoff.
We show the expected colors of star-forming 
galaxies at $4\la z \la 6$ with a red line, and mark
the colors for $z=4.0, 4.5, 5.0,$, and $5.5$ with open red
diamonds.  We show the colors of elliptical galaxies with
$0 \le z \la 2$ as a green line, and mark redshifts
$z=0, 0.5, 1,$ and $1.5$ with green triangles.  A dashed cyan
line marks the stellar
locus, based on Pickles (1998) templates, from M6 (top) to K1(bottom).
The dotted black line shows the selection region used by
Giavalisco et al (2004; see text).  The horizontal dashed line shows
the expanded region covered by our $V-i > 0.9$ cut for
inspection of spectra.  
\label{fig:colcol}}
\end{figure}

We compared our sample with the set of objects selected by 
the V dropout selection criteria that Giavalisco et al (2004)
developed for the GOODS project.  Those criteria are
\begin{eqnarray}
\nonumber && \left[ (V-i) > 1.5 + 0.9\times (i-z) ~~~ \hbox{\bf or} ~~~  
  (V-i)>2.0 \right]    \\
 && \hbox{\bf and} ~~~ (V-i)\ge 1.2  ~~~ \hbox{\bf and}
 ~~~ (i-z) \le 1.3 ~~. \label{giav_eqn}
\end{eqnarray}
We used the same $i<27.7$ magnitude cut when applying
equation~\ref{giav_eqn}.  Figure~\ref{fig:colcol} illustrates 
both selection criteria in the $(i-z, V-i)$ color-color plane.

\begin{figure}
\plotone{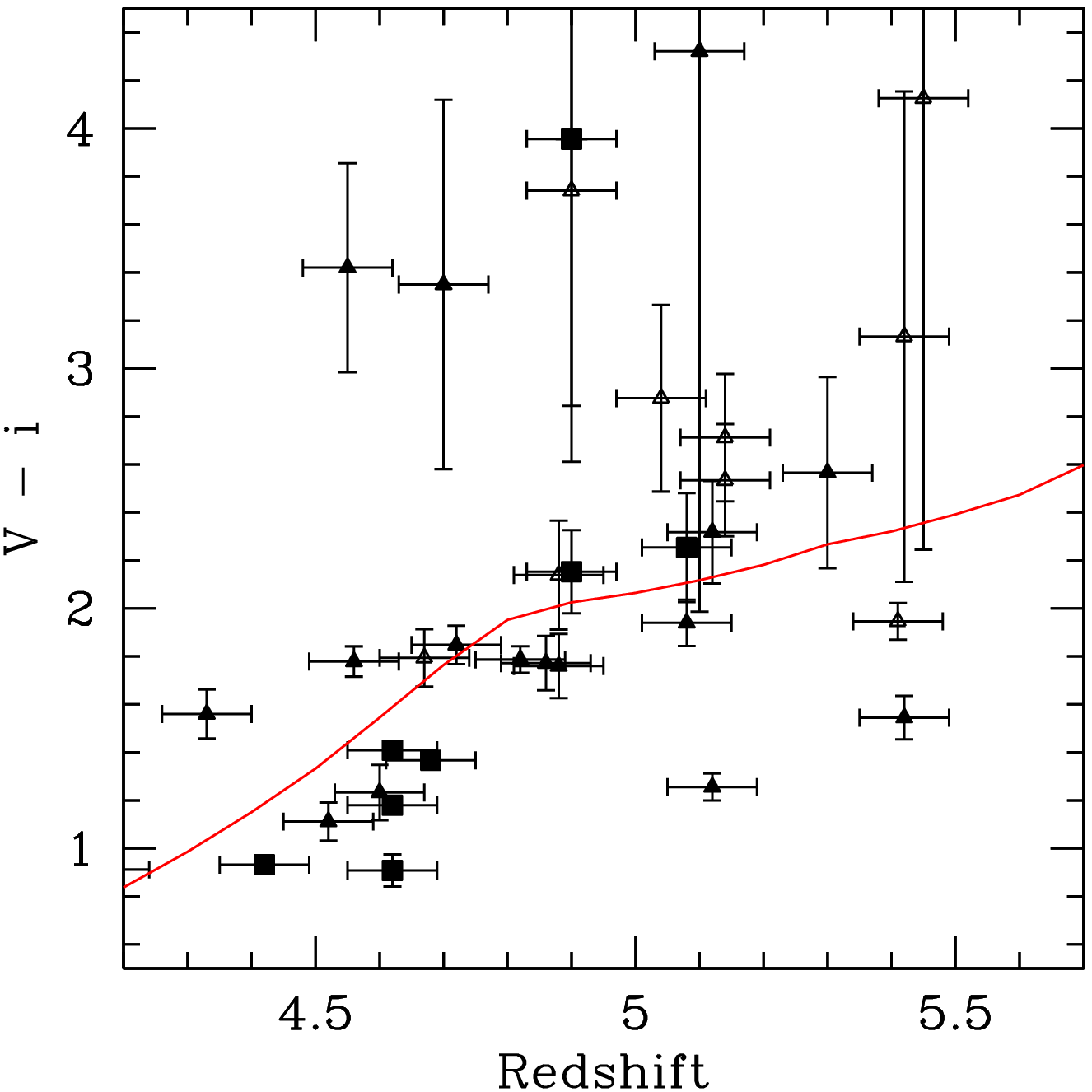}
\caption{The redshift dependence of $V-i$ color for spectroscopically
confirmed V-dropout Lyman break galaxies.  The expected color of a
template ``Magellanic'' spectrum is shown by the red curve.  The 
inflection point at $z=4.8$ corresponds to the redshift where the
\lya\ wavelength redshifts out of the V (F606W) filter and into the 
i (F775W) filter.  Points at $z < 4.8$ that fall blueward of this fiducial
line may in part be due to \lya\ emission in the V filter.
Point styles indicate the grade given to the Lyman break redshift
fit: Heavy squares are the best fits (assigned grade $<1.5$), 
filled triangles have grades between 1.5 and 2.1 (inclusive), and
open triangles have grades  between 2.1 and 2.5.
\label{viz}}
\end{figure}

Of the 39 good Lyman break galaxies, 25 meet the Giavalisco et al
selection criteria directly (eq.~\ref{giav_eqn}). One of these 
lies within $1\sigma$ of the selection
region boundary, as do five objects that would narrowly fail the 
Giavalisco et al criteria. 
Thus, the photometric V-dropout criteria 
have a completeness of $\approx 25/39 = 64\%$ compared to the
simple $V-i > 0.9$ cut of our spectroscopic sample.
Most of the confirmed V-dropout LBGs missed by eq.~\ref{giav_eqn} 
appear to be LBGs at a slightly lower redshift:  Our color-redshift
calculations indicate that a star forming galaxy should lie
at $z \ga 4.6$ to meet eq.~\ref{giav_eqn}, and $z \ga 4.3$ to
meet $V-i > 0.9$.  These redshift boundaries will of course
be blurred by variations in the galaxies' stellar populations,
dust content, and \lya\ line strength, all of which have some
effect on galaxy colors.  In practice, although only three of
our spectroscopically measured redshifts actually fall at $z<4.6$,
there is a reasonable correlation between $V-i$ and redshift ,
and the inclusion of objects with $0.9 < V-i \la 1.3$ does lower
the mean redshift of the sample (see figure~\ref{viz}).

The reliability of the photometric criteria is broadly comparable.
In total, we find 50 objects passing the criteria of eq.~\ref{giav_eqn}:
The 25 confirmed Lyman break galaxies discussed above, plus 27 other
sources {\it not} confirmed as Lyman break objects.  Among these, 21 had
inconclusive (grade 3) spectra, two are stars (grade-5), and only 
four are considered ``refuted'' with grade=4.  Leaving aside the
grade=3 sources, then, some
$\approx 25/31 = 81\%$ of the objects that meet the photometric
criteria are confirmed by inspection of their GRAPES/PEARS grism
spectra.  Combining this with the $64\%$ completeness would naively 
imply that V drop galaxy counts derived directly from eq.~\ref{giav_eqn} 
are slightly underestimated, by a factor of $0.64 / 0.81 \approx 0.8$.

The $i-z < 1.3$ color criterion in eq.~\ref{giav_eqn} 
seems unlikely to strongly affect which galaxies are included
in the sample.  The reddest galaxy actually selected
still has $i-z < 1$.  Thus, if there are redshift $z\sim 5$
Lyman break objects with $i-z>1.3$, they would have to form
a disjoint population from the star-forming sample discussed
in the present work.  This cut does, however, help exclude L and T
dwarfs from the sample.

A final point about sample selection is that Lyman
break galaxies constitute only $14 / 165 = 8\%$
of those objects {\it failing\/} the Giavalisco
et al criteria while still having $V-i > 0.9$ and $i < 27.7$.
Thus, for a purely photometric criterion, eq.~\ref{giav_eqn}
is quite good, and in the absence of spectra we could not
advocate a simple $V-i > 0.9$ selection.  
Expanding the selection region to include
objects with $i-z \la 0.2$ and $V-i > 0.9$ would 
be better than merely using $V-i > 0.9$ alone
(since the $i-z$ cut eliminates many interlopers
and no confirmed objects in our sample).  This would increase
the sample's completeness, but it would
still reduce the overall reliability of the photometric
selection when compared to the Giavaliso et al selection.

\begin{figure}
\plotone{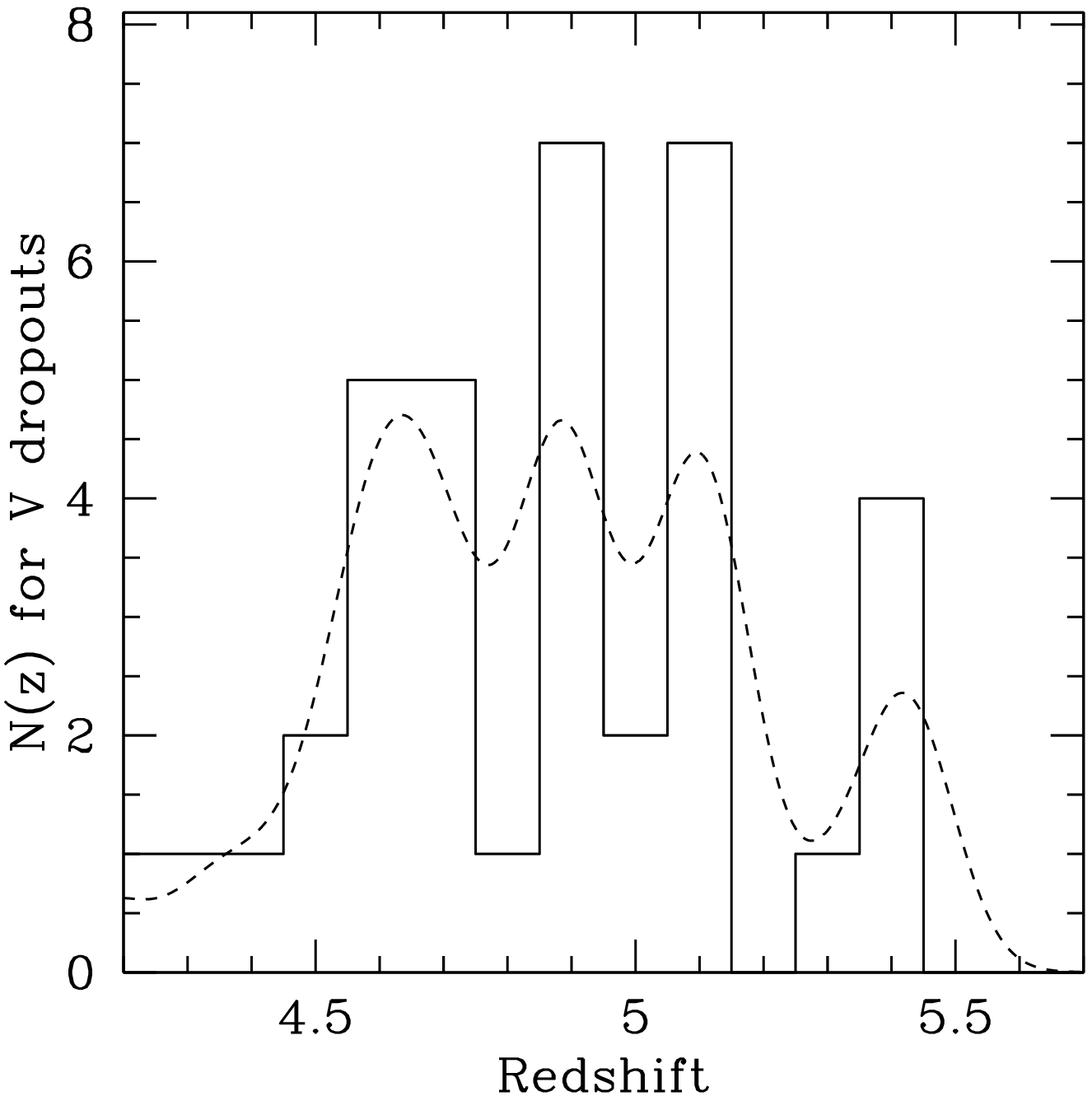}
\caption{{\it Solid:} Histogram of V dropout galaxy redshifts, based on
the $1 \le \hbox{grade} \le 2.5$ objects from table~\ref{vdrop_tab}.
{\it Dashed:} Generalized histogram of the same data, formed by
adding Gaussians centered at the redshift of each object
and having width $\sigma = 0.07$ (comparable to the redshift uncertainty
for a ``grade 2'' object).
The peaks in the distribution are not statistically significant.
\label{zhist}}
\end{figure}

We show the redshift distribution of the V dropout sample in
figure~\ref{zhist}.  The distribution shows no convincing structure
beyond a broad maximum.  In particular, we do not see any large scale
overdensity akin to the one we reported at $z\approx 5.9$ based on the
GRAPES $i$ dropout sample (Malhotra et al 2005).  The observed maximum
spans $4.5 \le z \le 5.2$, and can be understood by considering the
redshift-dependence of the selection criteria.  At $z<4.5$, the $V-i$
color is not red enough for the selection.  At $z>5.2$, reduction of
the $i$-band flux by the \lya\ forest is becoming significant.

We should expect the observed $V-i$ colors of the Lyman break objects to 
grow redder with increasing redshift, as the \lya\ forest absorption 
shifts through the V bandpass.  This effect is apparent in figure~\ref{viz},
though there is considerable scatter at each redshift.

\subsection{Lyman-$\alpha$ Emission Properties}
With spectra of every Lyman break galaxy in the Hubble Ultra Deep Field,
we can examine the statistics of \lya\ emission from galaxies in our
sample.  We have previously published lists of emission line
objects in the field (Xu et al 2007; Straughn et al 2008) and a more
specific study of the \lya\ galaxies, focussing on their ages,
masses, and morphologies (Pirzkal et al 2007; hereafter P07).  

Starting from the \lya\ galaxy sample of P07 and the V-dropout Lyman
break galaxy sample in this work, we find that five (ID numbers 712,
5183, 5225, 6139, and 9040) have previously identified \lya\ emission lines.
There
are an additional four \lya\ emitting galaxies in the P07 sample.
Three of these (ID numbers 631, 9340, and 9487) are too blue to be
included in our $V-i > 0.9$ sample, and the fourth (ID number 4442) is
too faint for our $i < 27.7$ criterion.  The P07 sample was
based essentially on an emission line search combined with a stringent
upper limit on $B$-band ($435$ nm) flux.  The completeness of the
emission line search is treated in Xu et al (2007).  That sample
is expected to be quite complete for  \lya\ galaxies that have 
$f_{Ly\alpha} \ga 2\times 10^{-17} \ergcm2s$, $EW \ga 120$\AA,
$4 \la z \la 6.5$, and $\theta_{FWHM} \la 0.5''$.  
We conclude that
$\sim 55\%$ of such \lya\ emitting galaxies also pass our V dropout
criteria.  While the remaining \lya\ galaxies are still well detected
in the photometric catalog, many fail the dropout color criteria.

In addition to these previously identified \lya\ objects, we find good
visual evidence for \lya\ emission in objects 1392, 2898, 3250, and 6515.
Weaker evidence of a possible line is seen in 
2408 and 8682.  Object 119 also shows a clear emission line, which 
could be \lya\ at $z=4.88$.  However, this line
was interpreted in Xu et al (2007) as \oiiipair.  (The ambiguity is
linked to the object's faintness, $i \approx 27.3$, and we
have assigned this object a grade of $2.5$ to reflect the 
uncertainty in its redshift.)  The detection of new \lya\ emitters
in the present paper is due to two factors.  First, we have
lower effective detection thresholds
in line luminosity and/or equivalent width in the current work, 
because our line list includes objects identified by visual inspection
of both 1-D and 2-D spectra in photometrically pre-selected high redshift 
galaxies.  Thus, an emission line does not need as high a statistical
significance to enter the present sample as was required in Xu et 
al (2007).  Second, we are using not only the GRAPES data but also
the PEARS-Deep data, nearly doubling the available data.
This provides better statistical signal-to-noise
ratio on most objects, and improved robustness to contamination by
overlapping spectra thanks to the additional four position angles of
data.

Combining all the detected \lya\ lines, we infer a \lya\ emission
fraction in the range $9/39 = 23\%$ to $12/39 = 31\%$ for the V dropout
sample.

There are two main ways in which HST grism searches can miss \lya\ emission
from well detected galaxies.  First, the emission line is always
located at the same wavelength as the continuum break introduced by
\lya\ forest absorption. This imposes a minimum observer-frame 
equivalent width that is comparable to the effective spectral 
resolution of the instrument.  This spectral resolution, in turn, 
is set by the angular size of the target, given the slitless 
instrument. 

Second,  \lya\ emission could come from a ``photosphere'' that is
more extended than the star-forming regions that dominate the UV
continuum light.  There is a plausible physical mechanism
for such extended \lya\ emission, namely, resonant scattering 
of \lya\ photons by neutral hydrogen in and around the emitting galaxy.
Moreover, observations of \lya\ ``blobs'' up to $\sim 10''$ in size
(e.g Steidel et al 2000) provide direct evidence that \lya\ can be scattered
or emitted over wide spatial scales, at least in some rare objects.
P07 report the effective radii in continuum emission for their
sample of nine \lya\ galaxies.  Among these, six have sizes
corresponding to $\la 1.5 \kpc$ FWHM, and the remaining three
have sizes near $3 \kpc$.
If the typical \lya\ photosphere is larger (i.e. $> 3 \kpc$ in extent), 
some \lya\ lines would be missed in the slitless HST spectra, because 
their large spatial extent would translate to very broad line widths.  $3
\kpc$ translates to $\sim 0.5''$, which in turn corresponds to a
400\AA\ observed line width.  Xu et al (2007) report a 70\% detection
completeness for 400\AA\ lines, but this completeness drops
(approximately linearly) to zero as the line width rises to 650\AA\ 
(corresponding to $0.8''$ size).

\begin{figure}
\plotone{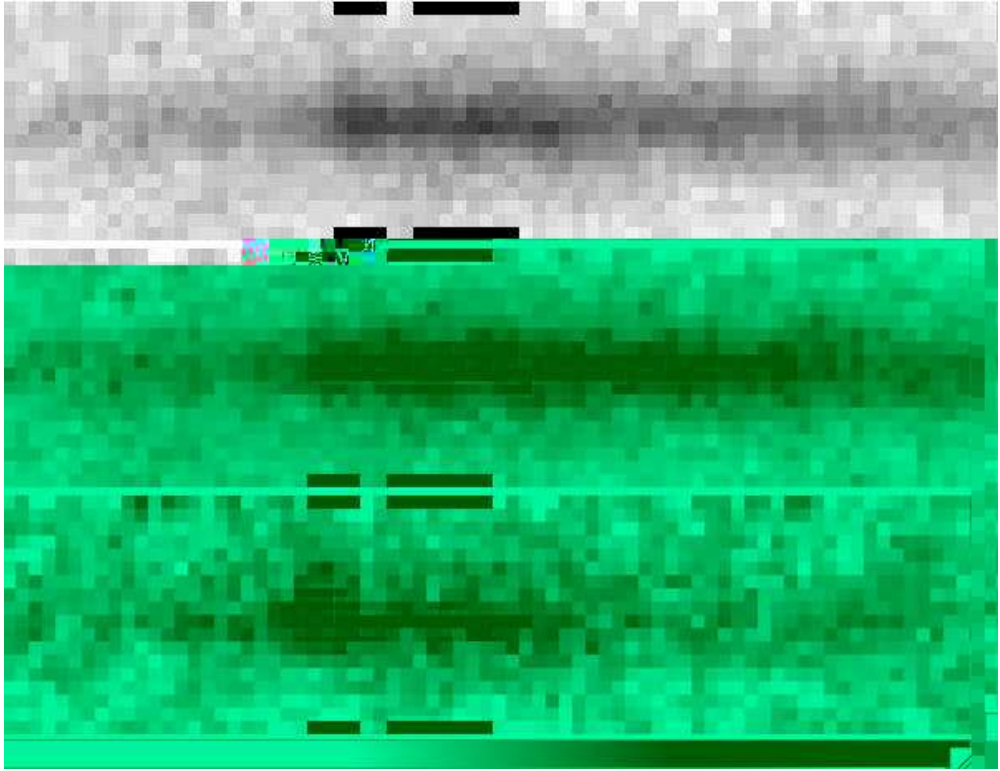}
\caption{2-dimensional stacks of V dropout grism data.  
{\it Top:} The full V dropout sample. {\it Middle:} Those
objects {\it without} individually detected \lya\ emission.
{\it Bottom:} Those objects {\it with\/} individually detected
\lya\ emission.
Each stack is performed using rectified, wavelength calibrated
grism stamps.  We used a ``shift and add'' algorithm.
We see a \lya\ line in the first and third stacked spectra, but 
no strong evidence for diffuse, extended \lya\ emission in the rest
of the population.  The pixel scale is 40\AA\ (observer
frame) in the spectral direction, and 0.05'' in the spatial direction.
The expected location of \lya, and comparison region of the
continuum spectrum, are marked by solid bars at the edges of
each spectrum.  These regions are used to produce the spatial
profiles in figure~\ref{fig:profcomp}.  Because there is
a range of redshifts in the sample and we have stacked the images
using observer frame wavelengths, there is no unique pixel - wavelength
correspondence away from 1215\AA.  We adopted this approach because the
observed \lya\ profile then corresponds directly to the size of
the \lya\ emitting region of the sample galaxies. 
The approximate total exposure times in the three stacks are 
$8.5\times 10^6$, $6.2\times 10^6$, and
$2.3\times 10^6$ seconds respectively.
\label{fig:stack2dv1}}
\end{figure}

\begin{figure}
\plotone{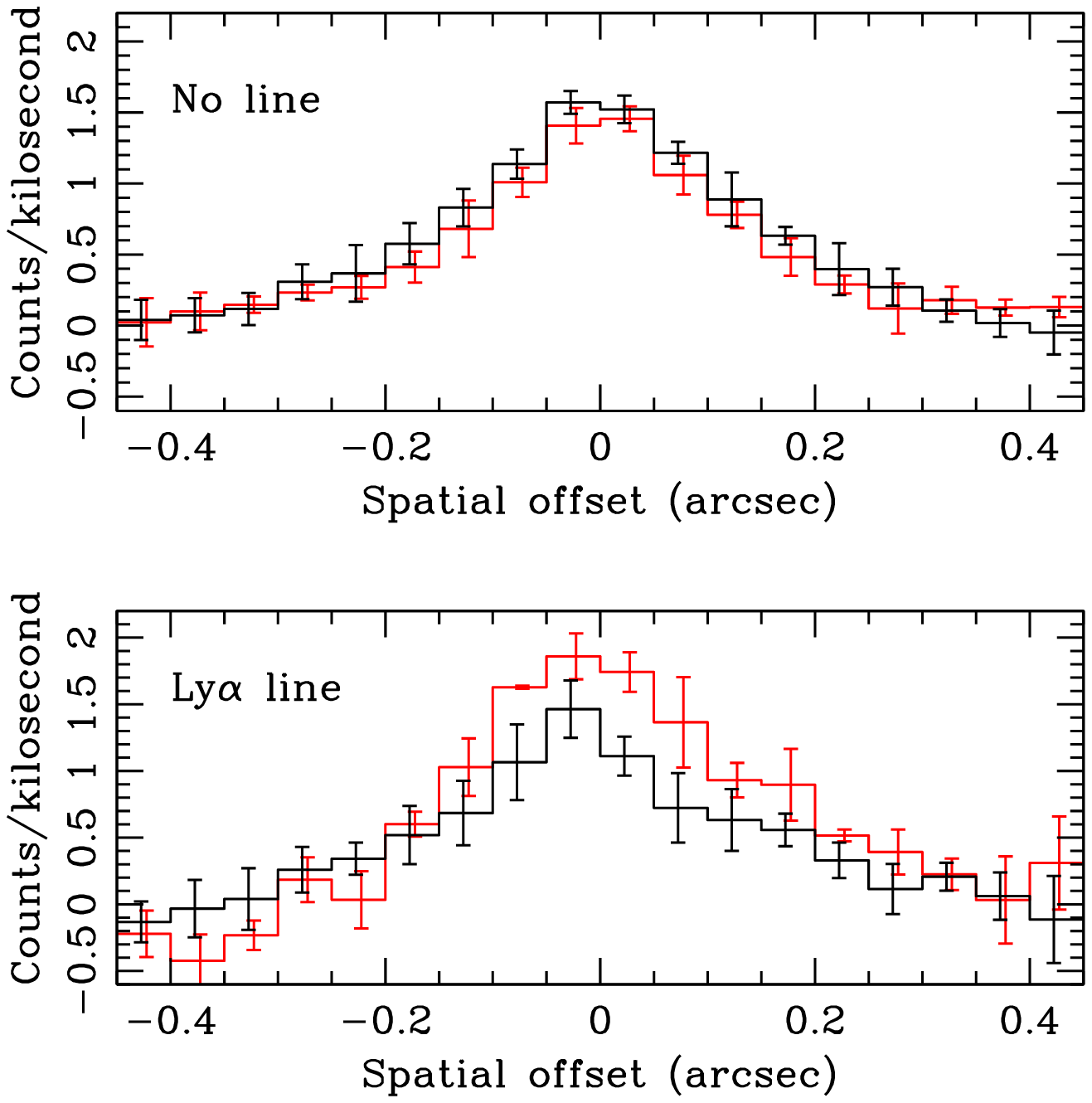}
\caption{Here we compare spatial profiles of the composite V-dropout 
spectrum (fig.~\ref{fig:stack2dv1})
at the \lya\ line wavelength (red line) and the adjacent continuum
at $\approx 1280$\AA\ rest frame (black line).  
The two profiles for the ``no line'' stack are indistinguishable within 
their combined $1\sigma$ error bars, implying that there is no 
substantial diffuse \lya\ component in this stacked spectrum.
The stack for the ``\lya\ line'' sample {\it does\/} suggest that
\lya\ comes from a modestly larger region than the starlight
in galaxies with prominent \lya\ lines, although the statistical
significance of the result is low with present data.
\label{fig:profcomp}}
\end{figure}

To address this possibility, we stacked the 2-dimensional grism
images of all 39 V-dropout galaxies, shifting them in wavelength
to align the expected location of any \lya\ emission in all spectra.
All position angles were included in the average.
The resulting average 2D spectrum is shown in figure~\ref{fig:stack2dv1}.
If spatially extended \lya\ flux is prevalent in our sample, we should
expect to see a region of diffuse emission at the \lya\ wavelength.
Moreover, the profile of this emission would be a fair representation
of the ``average'' \lya\ emission morphology.
In practice, we see no evidence for extended \lya\ emission in this
figure.  The continuum emission, a \lya\ break, and weak transmitted
flux through the \lya\ forest region are all clearly detected.  
We also made a stack of the ten best \lya\ emitters, and a stack
of all the remaining 29 objects.  The ``\lya stack'' shows a clear
emission line.  Moreover, this line appears somewhat extended 
on visual inspection (figures~\ref{fig:stack2dv1} and~\ref{fig:profcomp}).
The difference is not highly significant, but a direct measurement
of the spatial width of the spectrum yields
FWHM=0.19'' (or $\approx 1.3 \kpc$) in the continuum region, 
and 0.26'' (or $\approx 1.8 \kpc$) in the line region.

As a quantitative test for extended \lya, we have compared the 
spatial profile of the stacked spectrum (perpendicular to the
dispersion direction) at both the \lya\ location and in the
adjacent continuum region.  We find that the two profiles are
indistinguishable (figure~\ref{fig:profcomp}) .  
Thus, while we cannot rule out extended \lya\ 
emission in a minority of cases, it is at most an exception to the
rule, rather than a generic phenomenon in Lyman break galaxies.
To place a quantitative upper limit on the flux in a spatially
extended component, we first measure the $1\sigma$ noise level 
in the composite spectrum, which 
is about $1/3$ count/ksec/pixel.  This corresponds to a flux
level of $2\times 10^{-18} \ergcm2s$ in one $0.05''$ ACS WFC pixel.
If we co-add the results for a larger solid angle $\Omega$, 
we expect the limit to scale as $\sqrt{\Omega / \Omega_{pix}} = 
\sqrt{N_{pix}}$.  We define two sets of pixels, each consisting of
two $1.05'' \times 0.3''$ rectangular regions and so totaling 
$0.63 \sqarcsec$, and each excluding a strip $0.3''$ wide 
along the trace of the stacked spectrum.
The first set of pixels is centered at the expected location 
of \lya\ emission, while the second is centered $1.05''$ away,
towards the red end of the spectrum.  We take the total
fluxes in these two sets of pixels, apply the wavelength-dependent
sensitivity conversion factor, and compare the results.  In the
region where diffuse \lya\ might be expected, we find a formal
flux excess of $(5 \pm 4) \times 10^{-17} \ergcm2s$.
This is comparable to the typical fluxes of ground-based narrowband \lya\ 
surveys.

\section{Summary}
We have examined GRAPES and PEARS spectra of 216 photometrically selected
candidate Lyman break galaxies, and spectroscopically confirmed 39 
of them.   Our pre-selection used a wider range of color space ($V-i > 0.9$)
than more traditional ``V-dropout'' Lyman break color criteria, in
order to assess the completeness of those criteria, along
with their reliability.  We find that 64\% of our confirmed 
objects meet the traditional criteria.  Among those galaxies
passing the traditional criteria, and having adequate
spectra for classification, we find that $81\%$ are 
confirmed as Lyman break galaxies.  
Our V drop sample includes 55\% of \lya\ emitting galaxies previously 
identified in this redshift range using GRAPES data (Xu et al 2007,
Pirzkal et al 2007).  The ``missing'' \lya\ galaxies are either
too blue in $V-i$ (in part due to the emission line in the V filter)
or too faint in the continuum for our V dropout selection criteria.
We also detect 4--7 additional \lya\ galaxies not included in
the earlier samples.  
Our overall \lya\ detection fraction is comparable to that
in spectroscopic followup of other Lyman break surveys (e.g., Steidel
et al 2000).
We have examined our stacked 2D spectra for
evidence of the diffuse \lya\ emission that might result from
resonant scattering of \lya\ photons in neutral hydrogen near
young galaxies.  In the overall stack of 39 galaxies, we find 
no significant evidence for such emission down to a  $2\sigma$ 
upper limit of $9 \times 10^{-17} \ergcm2s$ over an $0.6\sqarcsec$ 
region.
On the other hand, when we examine the composite spectrum of
just those galaxies with individually identified \lya\ lines,
we see a modestly broader spatial profile at the wavelength
of \lya\ than in the adjacent continuum.  This is consistent
with the possibility that scattering of \lya\ photons results
in a somewhat extended \lya\ photosphere in these objects.

\acknowledgements
This work has been supported under grant numbers HST-GO-09793
and HST-GO-10530.  C.G. acknowledges support from NSF-AST-0137927.
L.A.M. acknowledges support by NASA through contract number 1224666
issued by the Jet Propulsion Laboratory, California Institute of
Technology under NASA contract 1407.
We thank the STScI staff in general, and Beth Perriello in particular,
for their assistance with the planning, scheduling, and execution of the 
GRAPES and PEARS programs.

\end{document}